**Mechanistic insight into the limiting factors of graphene-based environmental sensors**

*Jean-Michel Guay, Ranjana Rautela, Samantha Scarfe, Petr Lazar,Saied Azimi, Cedric Grenapin, Alexei Halpin, Weixiang Wang, Lukasz Andrzejewski, Ryan Plumadore, Jeongwon Park, Michal Otyepka, Jean-Michel Ménard, and Adina Luican-Mayer\**

Dr. J.-M. Guay, Dr. R. Rautela, S. Scarfe, C. Grenapin, Dr. A. Halpin, W. Wang, Lukasz Andrzejewski, R. Plumadore, Prof. J.-M. Ménard, Prof. A. Luican-Mayer
Department of Physics, University of Ottawa, Ottawa, Ontario, K1N 9A7, Canada
E-mail: luican-mayer@uottawa.ca

S. Azimi, Prof. J. Park
School of Electrical Engineering and Computer Science, University of Ottawa, Ottawa, Ontario, K1N 6N5, Canada

Petr Lazar, Michal Otyepka
Regional Centre of Advanced Technologies and Materials, Faculty of Science, Palacky University Olomouc, Slechtitelu 27, 771 46 Olomouc, Czech Rep.

Graphene has demonstrated great promise for technological use, yet control over material growth and understanding of how material imperfections affect the performance of devices are challenges that hamper the development of applications. In this work we reveal new insight into the connections between the performance of the graphene devices as environmental sensors and the microscopic details of the interactions at the sensing surface. Specifically, we monitor changes in the resistance of the chemical-vapour deposition grown graphene devices as exposed to different concentrations of ethanol. We perform thermal surface treatments after the devices are fabricated, use scanning probe microscopy to visualize their effects on the graphene sensing surface down to nanometer scale and correlate them with the measured performance of the device as an ethanol sensor. Our observations are compared to theoretical calculations of charge transfers between molecules and the graphene surface. We find that, although often overlooked, the surface cleanliness after device fabrication is responsible for the device performance and reliability. These results further our understanding of the mechanisms of sensing in graphene-based environmental sensors and pave the way to optimizing such devices, especially for their miniaturization, as with decrising size of the active zone the potential role of contaminants will rise.

# 1. Introduction

The thinnest possible material, graphene, and other atomically thin inorganic graphene-like materials are thought to have great potential for next-generation flexible electronics and optoelectronic devices because of their mechanical strength and flexibility, high electron mobilities that lead to low power consumption, and unprecedented properties enabled by confining electrons in low dimensions.[1] For environmental gas monitors, two-dimensional (2D) materials have already demonstrated promising performance for figures of merit such as: sensitivity, stability, and response time.[2,3] Graphene represents an ideal sensing platform having large surface area, chemical stability and enormous sensitivity to changes in its environment. It has high density and mobility of charge carriers, low electrical noise and it enables sensing of both electron acceptor and donor molecules [4,5]. However, the development of sensing applications is still hampered by the lack of precise understanding of the mechanisms underlying the sensing response to the chemical and biological species present on the surface of 2D materials [6]. This is an especially complex question as the sensor's sensitivity, response time, and reversibility are highly dependent on the graphene synthesis as well as device preparation methods. Reliabale processes of fabrication and sensor "pretreatment" are yet to be established. Most of the reported literature takes a trial and error approach to exploring sensing properties, resulting in an acute need for deeper understanding of the sensing mechanisms and how they can influence the device performance. Currently, sensors based on large-scale high-quality graphene growth methods such as chemical vapor deposition (CVD) are less explored than chemical derivatives of graphene[7-9].

Here, we address the need for in-depth inquiry into the sensing mechanisms by studying devices based on CVD-grown graphene as they are exposed to one of the most common volatile organic compounds (VOC), ethanol. We explore the consequences of distinct surface treatment procedures and device geometries and interpret them by corroborating microscopic surface studies obtained by scanning probe microscopy and theoretical calculations of molecule/surface interactions. We find that, although often overlooked, small residues at the surface of graphene dramatically affect its electrical properties in gas sensing applications. Our work also investigates, for the first time, the potential of CVD graphene-based devices for ethanol vapour detection and proposes optimal routes for sample fabrication leading to an improved reliability.

## 2. Results and Discussion

### 2.1. Sensor Characteristics/Performance

Our specific design for devices is presented in the schematic of Figure 1a and the pictures of electrically connected devices in Figure 1b. The exact fabrication methods are detailed in the experimental section. Graphene was transferred to cleaved $Si/SiO_2$ wafers and patterned into a device using a shadow mask for Au evaporation. A schematic of the set-up used to test the response of the sensors is shown in Figure 1c.

In this experiment, we focused on sensors fabricated using graphene produced by chemical vapor deposition (CVD)[8, 9]. Although these films are among the leading contenders for industrial applications, CVD methods do not result in perfectly crystalline graphene sheets.[10] Therefore, their properties might differ from what is expected for intrinsic graphene [11-13]. Most common defects, as schematically presented in Figure 1d, include missing atoms, Stone-Wales defects, and extended grain boundaries, areas with more than a single layer of graphene, folds, and wrinkles.[14-17] As these defects are characterized by specific electronic states localized at the nanoscale [15], their presence will inadvertently affect the electrical response of the CVD graphene upon exposure to airborne chemical species. The precise control and characterization of nano-scale defects is therefore crucial to improve performances and reliability of these promising environmental sensors.

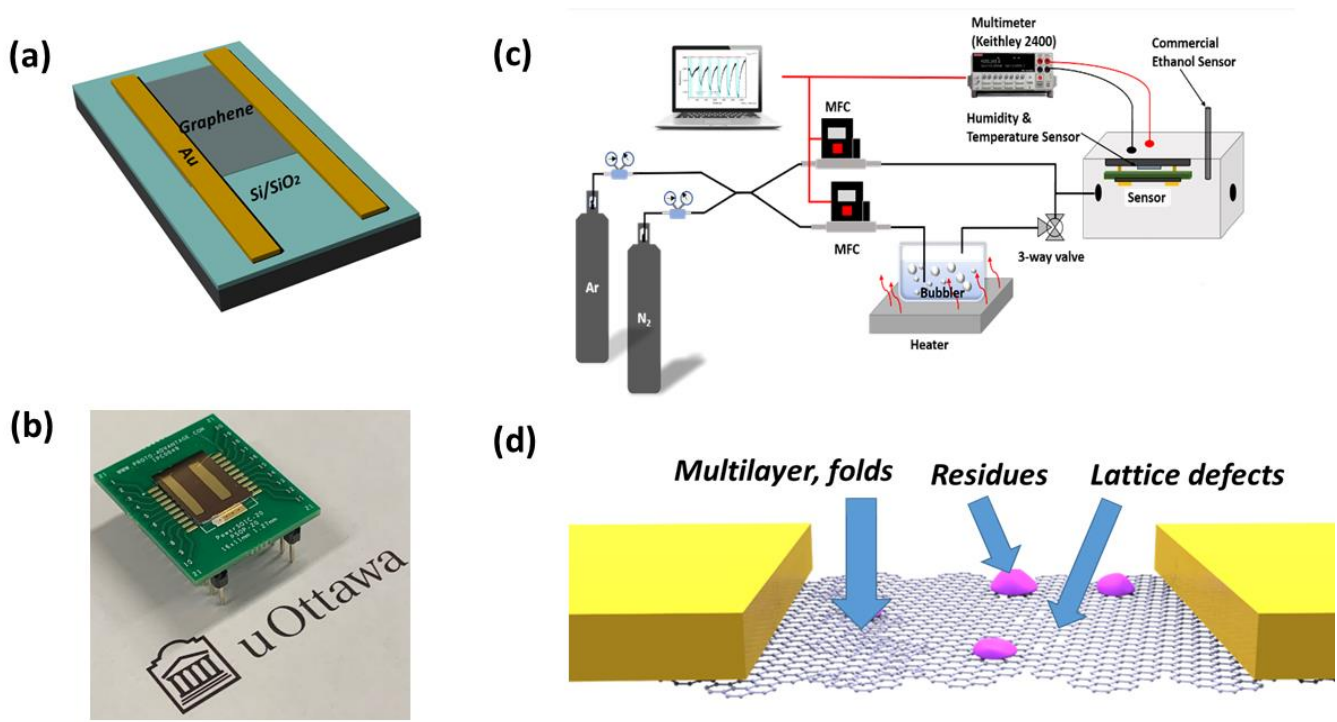

**Figure 1.** Sensor design and test set-up. (a) Schematic of sensor design (b) Picture of the sensor geometry (c) Schematic of experimental setup to test the graphene sensors (MFC – mass flow controller). (d) Schematic of the sensing surface and its imperfections.

A factor that strongly influences graphene's surface is also related to the process of transfer from the Cu foil onto a wafer, which is a necessary process for any device fabrication. The need to place protective polymers such as polymethylmethacrylate (PMMA) on top of graphene during the etching of the copper foil or during the lithographic processes inadvertently results in the presence of polymer residue on the surface of graphene as suggested in Figure 1d. Here, we address this limiting factor by controlling the amount of residue and measure its influence on the sensor performance.

Generally, immersion in acetone is the first step performed to remove PMMA layer on the transferred CVD graphene. However, this inevitably leaves contaminants on the surface of graphene due the high molecular weight of PMMA [6]. Thermal treatment or annealing in a forming gas environment is also known to result in reducing the amount of polymer residue [13].

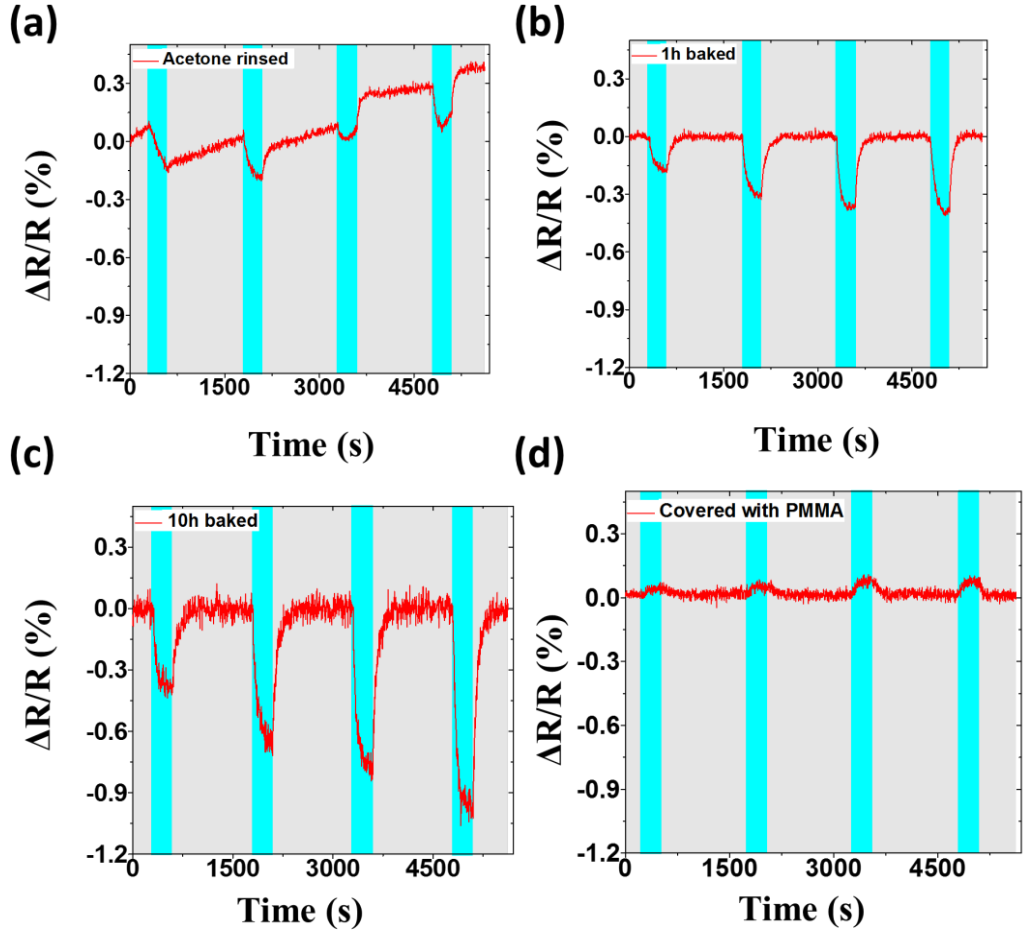

**Figure 2.** Evolution of resistance as a function of time for graphene sensors that have had the following surface treatments (a) rinsing in acetone (b) baking in forming gas for 1h (c) baking in forming gas for 10h. (d) Graphene sensor covered with 60 nm PMMA. The sensors were exposed to concentrations of 1400, 2730, 4000, and 5200 ppm, respectively in this order. We note that a baseline was subtracted from the data for clarity.

Figure 2 presents the summary of graphene sensors responses upon various surface treatments. The sensors were exposed successive concentrations of 1400, 2730, 4000, and 5200 ppm of ethanol, respectively. In between each exposure of the sensors to ethanol, the chamber was purged with $N_2$ for a period of 20 minutes. The concentration of the ethanol was obtained using Antoine's equation and the proper parameters from partial pressure tables for ethanol.[18]

Figure 2a shows an example of sensing response for a device that was cleaned by wet removal of PMMA in acetone (see methods). We find that sensors treated using this method have low reliability and large device-to-device variations. An example of response of sensors that had an additional step of one hour baking, further removing PMMA residue as discussed below, is show in Figure 2b. In this case, we observe improvement over device reliability and sensitivity. When the sensor surfaces are baked for 10 hours (Figure 2c), however, they are superior both in terms of sensitivity and reliability, showing little device-to-device variations.

We also analyzed the case when the sensor was fully covered by PMMA (Figure 2d) and the response was found to be very weak (below 0.1% sensitivity) and had the opposite sign (the resistance increases upon exposure) demonstrating that graphene (and not PMMA) is the key participant in the mechanisms responsible for sensors with higher sensitivity. Previous reports for different graphene sensor geometries [19,20] suggested that devices that have not been baked showed highest sensitivity. In contrast, we find that the more PMMA residue on the surface, the stronger the device-to-device variations.

We will now present our understanding of these results from corroborating detailed surface studies of graphene with theoretical Density Functional Theory (DFT) calculations.

PMMA residues lead to an intrinsic electrical doping concentration, inherently modifying the sheet resistance of the graphene sensors. [21] To remove contaminants, Ishigami et al. showed that graphene could be oven-baked in a mixture of $Ar/H_2$ gas.[22] The effect of baking the surface of graphene in forming gas can be readily seen in Figure 3a, where from left to right we present atomic force microscopy of regions on a graphene sensor that was rinsed in acetone without baking, baked at 400 °C for 60 minutes, and for 9 hours, respectively. As noticed in these topographic images, the cleaning process was not 100% efficient at removing contaminants from the surface. The contaminants are believed to be residue of the PMMA used during the transfer of graphene onto the $Si/SiO_2$ substrate. The size of the contaminants is also observed to increase during the baking process. Statistical analysis of the surface shows a reduction of residue coverage from 25% to 8% by baking and clustering of the residue, resulting in measured dimensions of the contaminant regions from 4 nm to 95 nm. The average roughness was observed to significantly decrease with baking.

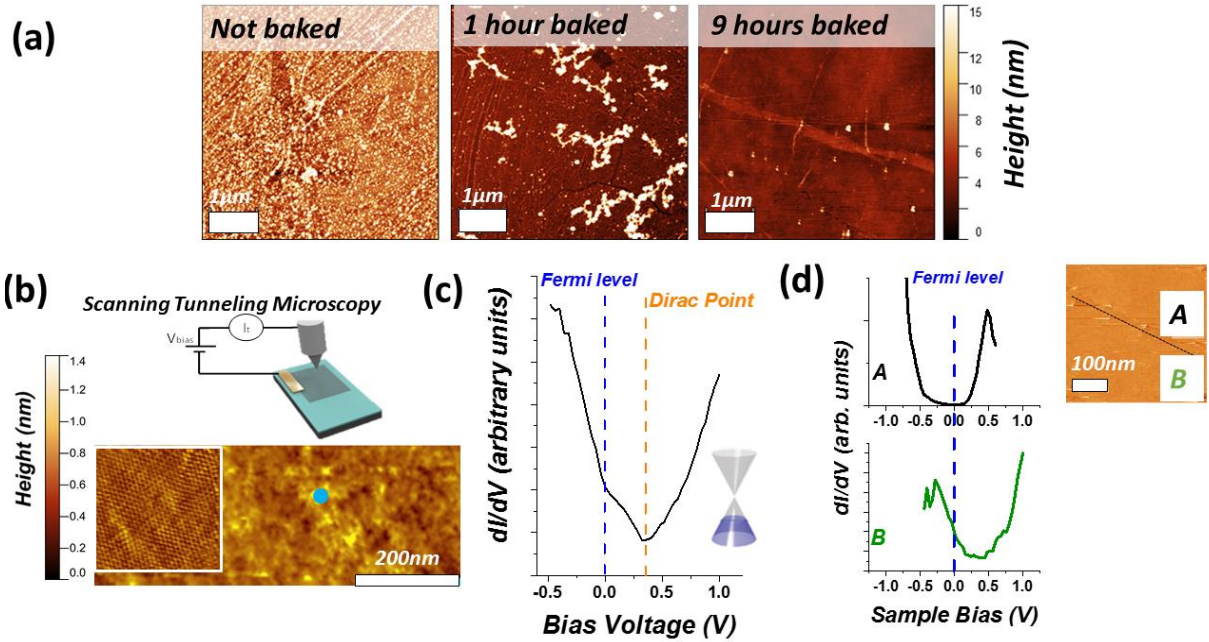

**Figure 3.** Microscopic characterization of the sensing surface. (a) Atomic force microscopy images of the sensing surface after different baking time at 400 °C. (b) Top: Schematic of the STM experiment; Bottom: STM topographic image with inset of 6 nm × 6 nm atomically resolved lattice. The blue dots indicates the position of the spectrum in (c). (c) STS on the blue dot in (b) together with pointers to the Fermi level and Dirac Point. The inset cone suggests p-doping. (d) STS of two different adjacent regions as indicated. Inset: STM topographic image showing the areas where the data in (d) was taken.

After sufficiently long time, approximately 9 hours, the surface of graphene is mostly free of residue, with less than 1% surface covered by PMMA contaminants. In that situation, the atomically clean lattice was accessible for scanning tunneling microscopy and spectroscopy experiments, at room temperature, to locally characterize the morphological and electronic properties of the sensing surface. Large areas are found to be free of residues and of up to 1 nm roughness, as the graphene conforms to the $SiO_2$ surface (Figure 3b). When we zoom in, we can image the lattice of graphene with atomic resolution, as presented in the inset of Figure 3b.

By recording the tunneling differential conductance, dI/dV, scanning tunneling spectroscopy gives access to the local density of electronic states. The spectrum recorded at the point indicated in Figure 3b is presented in Figure 3c. We indicate the position of the Fermi level and Dirac point of graphene. In neutral graphene, these two points coincide [23]. Here, we find that, although locally clean, the graphene sensor has a p-doping level $n = 8 \times 10^{12} cm^{-2}$, corresponding to a shift in Fermi level of ~330 meV.

Interestingly, while the p-doping is a representative situation, another type of region we encountered less frequently is presented in Figure 3d. We present the spectroscopic data acquired in two adjacent regions (Areas A and B), separated by an extended defect (dashed, black line), together with the topographic image in the inset. In this case, the two regions show different doping level, where Area A is less doped. One possible explanation could be the effect of the SiO$_2$ substrate that is known to host an electrostatic potential due to the presence of randomly distributed trapped charges in the oxide. Another possible scenario is the presence of a fold in the graphene surface, which is a common occurrence at the transfer step. A fold would result in a multilayer region, with more metallic electronic band character than a monolayer, and thus with a Fermi level less affected by the presence of doping. This further underscores the important role played by nanometer scale surface imperfections.

To discuss the sensor response, we now focus on the sensing mechanism. Generally, as graphene is charged away from the neutrality point by adsorbates that act as either acceptors or donors, its resistance changes. We schematically present in Figure 4a the basics of the ambipolar field effect in graphene [24]. To understand the effect of a particular airborne chemical, we must first know the doping level of the graphene sensors before exposure to the gas. In this case, from the STM data discussed above as well as the gating response measured on similar devices, we found the graphene sensors prepared by our methods to be p-doped. To explore the sensing mechanism in the system ethanol/graphene, we performed Density Functional Theory (DFT) calculations as detailed in the method section.

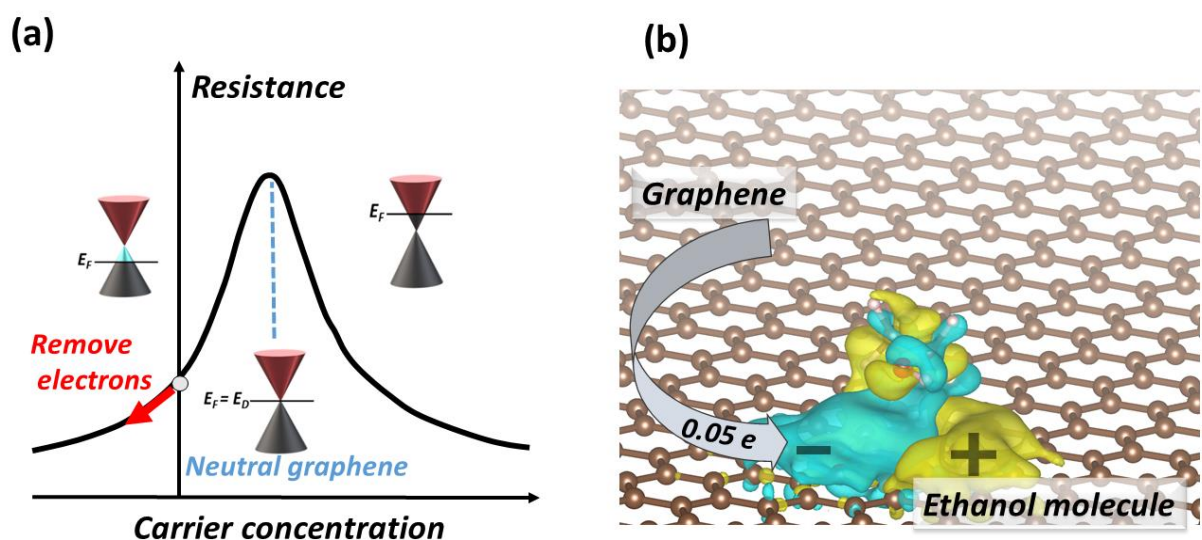

**Figure 4.** Sensing mechanism. (a) Schematic of the effect that carrier concentration has on graphene's resistance. (b) The geometry of adsorbed ethanol molecule and the charge density distribution. The colors indicate negative (blue) or positive (yellow) charge difference

In Figure 4b we plot the geometry of adsorbed ethanol molecule and the charge density difference induced by assembling the system from isolated parts (ethanol molecule and graphene sheet). The colors indicate negative (blue) or positive (yellow) charge difference. This illustrates that there is indeed charge redistribution between ethanol and graphene which is the mechanism responsible for sensing. Moreover, the plot also shows that although the net charge transfer is such that the ethanol molecule acts as an acceptor and removes electrons from graphene, the charge transfer process is more complex. Upon adsorption, the molecule appears to locally polarize graphene and both positive and negative charge carriers are introduced. The calculated charge transfer is $0.05e$ from graphene to a single ethanol molecule. In a realistic situation, however, the ethanol molecules tend to aggregate into hydrogen-bonded clusters (dimers, tetramers, pentamers) at higher surface coverage of ethanol on the hydrophobic graphene surface.[25] In that case, we calculate the charge transfer from graphene to ethanol is $0.048e$, $0.025e$, and $0.015e$ (per molecule) for single ethanol, ethanol dimer, and ethanol tetramer, respectively. The ethanol clusters still behave as p-dopants but the clustering of ethanol molecules decreases p-doping of graphene per ethanol molecule, which is consistent with the observation of lowering of sensor response with increasing ethanol concentration.

In our calculations, we also find that PMMA acts as an acceptor, which is consistent with the observation that typically devices that have residues are p-doped.

When PMMA is involved in the sensing mechanism, however, the process is more complex and is likely responsible for the large device-to-device variations in sensors with large amounts of polymer residues.

## 3. Conclusion

In conclusion, by combining sensing experiments with surface probe techniques and theoretical calculations, we provide insights into the sensing mechanisms at the surface of CVD graphene when exposed to ethanol environments.

We find that, the degree to which the sample still has polymer residues from the fabrication process dominates the sensing mechanism. Specifically we find that in the high polymer coverage regime, there is large device-to-device variations. In contrast, the samples with least polymer residue have high reproducibility and show superior sensitivity. This results highlight that, although often discounted, the interplay between PMMA residues, the sensing surface and its defects, play a significant role in its response. Thus, we provide a guide for optimizing graphene-based sensors, suggesting that in order to achieve a reliable sensor based on CVD graphene, full control over contamination should be established.

## 4. Experimental Section

*Sensor Fabrication:* For this experiment we used commercially available graphene grown by chemical vapour deposition (CVD) on copper sheets (G/Cu-60-40, Graphenea). Square sections of $1 cm \times 1 cm$ of graphene/copper foil were covered with a 500 nm layer of polymethyl-methacrylate (PMMA). The back side of the foils were oxygen etched (RIE-10NR, SAMCO) at 100 W for 1 minute to remove any graphene residue, to ensure proper graphene transfer and complete etching of the copper foil. The samples were then placed into copper etchant (667528, Sigma-Aldrich) heated at a temperature of 70 °C. After the complete removal of the copper foil, the PMMA/graphene were removed from the solution and placed on $Si/SiO_2$ substrates (HS39626, Nova Electronic Materials). Prior to transfer, the $Si/SiO_2$ substrates were rinsed in acetone, sonicated for 5 minutes, rinsed with IPA, dried with $N_2$ and baked in air at 180 °C for 5 minutes.

*Graphene sensor*: Graphene transferred to cleaved $Si/SiO_2$ wafers were placed and mounted in a Thermo/Electron Beam Evaporator (Nexdep Series, Angstrom). A shadow mask was placed in front of each the sensors during the deposition of the two metallic electrodes: 5 nm Ti/ 150 nm Au. Following the deposition, the samples were rinsed in acetone, IPA and dried with $N_2$.

*Testing Apparatus:* High purity grade argon (or nitrogen) gas was used as the carrier and mixing gas during the experiments. The gas was directed into two lines. The first line consisting of a mass flow controller MFC (0-1000 sccm) (MC-1000SCCM, Alicat Scientific) to purge the chamber and control the concentration of ethanol. The second line composed of an MFC (0-100 sccm) (MC-100SCCM, Alicat Scientific) connected to a bubbler (31-500-702, Fischer Scientific) for carrying the ethanol vapors to the test chamber. The bubbler was left at room temperature and monitored via a thermocouple. A manual 3-way valve was used to divert the ethanol vapors into the fume-hood during the purging of the chamber. A calibrated commercial ethanol sensor (ETH-BTA, Vernier) was mounted onto the chamber and used as reference to quantify the ppm levels of ethanol inside the testing chamber. A temperature and humidity sensor (1649-1012-2-ND, Digi-Key) were placed under the graphene sensor as well as in the flow to monitor the humidity and temperature condition at the surface of graphene. The resistance was obtained from the linear I/V characteristic of each sensor.

*Theoretical calculations*: Density Functional Theory (DFT) calculations were performed using the projector-augmented wave method in the Vienna Ab initio Simulation Package (VASP) suite.[26,27] The optB86b-vdW DFT functional was employed to include a contribution from non-local correlation effects.[28,29] The graphene sheet was modeled using a 6×6 supercell (72 carbon atoms) with a calculated C-C bond length of 1.44 Å. The periodically repeated single-layers were separated by at least 18 Å of vacuum. The energy cutoff for the plane-wave expansion was set to 400 eV and 3x3x1 k-point grid was used. The charge transfer was calculated using iterative Hirshfeld (Hirshfeld-I) charge partitioning. In the iterative Hirshfeld algorithm, the neutral reference atoms are replaced with ions with fractional charges determined together with the atomic charge densities in an iterative procedure.[30] It should be noted that the size of the charge transfer is to some extent dependent on the method used to calculate it. We checked to results using the classical Hirshfeld charge analysis, and resulting charge partitioning was similar on both magnitude and direction.

**Acknowledgements**

A L-M, J-M M, J P Acknowledge funding through DND Ideas, A L-M acknowledges funding from NSERC Discovery, NSERC SPG – P. J-M M acknowledges funding from NSERC. We acknowledge the support of CMC Microsystems.

P L, M O acknowledge support by the Operational Programme Research, Development and Education-European Regional Development Fund, project no. CZ.02.1.01/0.0/0.0/16_019/0000754 of the Ministry of Education, Youth and Sports of the Czech Republic and by an ERC Consolidator Grant (H2020) No. 683024).